\documentclass[letters,usenatbib]{mnras}
\usepackage{ulem}
\usepackage{amsmath}
\usepackage{graphicx}
\usepackage{subfigure}
\usepackage{multirow}
\usepackage{color}
\usepackage{enumitem}
\usepackage[dvipsnames]{xcolor}
\begin{document}
\title[$\gamma$-Rays in Arp 220]{$\gamma$-Ray Emission from Arp 220: Indications of an Active Galactic Nucleus}

\author[Yoast-Hull et al.]{Tova M. Yoast-Hull$^{1}$\thanks{E-mail: yoasthull@cita.utoronto.edu}, John S. Gallagher III$^2$, Susanne Aalto$^3$, and Eskil Varenius$^3$\\
$^1$Canadian Institute for Theoretical Astrophysics, University of Toronto, ON, M5S 3H8, Canada\\
$^2$Department of Astronomy, University of Wisconsin-Madison, WI 53706, USA\\
$^3$Department of Earth and Space Sciences, Chalmers University of Technology, Onsala Observatory, 439 92 Onsala, Sweden}

\maketitle


\begin{abstract}
Extragalactic cosmic ray populations are important diagnostic tools for tracking the distribution of energy in nuclei and for distinguishing between activity powered by star formation versus active galactic nuclei (AGNs).  Here, we compare different diagnostics of the cosmic ray populations of the nuclei of Arp 220 based on radio synchrotron observations and the recent $\gamma$-ray detection.  We find the $\gamma$-ray and radio emission to be incompatible; a joint solution requires at minimum a factor of $4 - 8$ times more energy coming from supernovae and a factor of $40 - 70$ more mass in molecular gas than is observed.  We conclude that this excess of $\gamma$-ray flux in comparison to all other diagnostics of star-forming activity indicates that there is an AGN present that is providing the extra cosmic rays, likely in the western nucleus.
\end{abstract}


\section{Introduction}

Distinguishing between nuclear activity powered by starbursts versus active galactic nuclei (AGNs) is often a difficult task, particularly in systems which are heavily obscured by dust \citep[e.g.,][]{Teng12}. The combination of star formation and dense gas is especially problematic in the nuclei of major mergers that are possible sites for the significant growth of super-massive black holes (SMBHs) during the present epoch via the radiative mode \citep[e.g.,][and references therein]{Kormendy13,Heckman14}. The dense interstellar medium (ISM) which supports radiative accretion may also fuel luminous starbursts whose presence masks signatures of the AGN. Furthermore, the circumnuclear ISM can be sufficiently dense to be Compton-thick, thereby obscuring diagnostic X-rays and mid-infrared emission from an AGN.  

In the case of the nearby ultra-luminous infrared galaxy (ULIRG) Arp 220, the ISM in the two starburst nuclei is notoriously dense with optical depths greater than 1 out to 860 $\mu$m \citep[e.g.,][]{Downes07,Sakamoto08,Wilson14}.  Indirect arguments suggest that an AGN produces a significant fraction of the radiated power in the western nucleus \citep{Wilson14,Rangwala15,Martin16}.  However, despite numerous observations from submillimeter to meter wavelengths \citep[e.g.,][]{Sakamoto08,Salter08,Aalto15,Barcos15,Scoville15,Martin16,Varenius16} and in both soft and hard X-rays \citep{Paggi13,Teng15}, the presence of an AGN in one or both nuclei has yet to be firmly established.

The recent \textit{Fermi} detection of Arp~220 in high-energy $\gamma$-rays \citep{Griffin16,Peng16} provides unprecedented insight into the energetics of the system.  The $\gamma$-ray flux keeps pace with the total infrared luminosity of Arp 220 such that the galaxy lies on the emerging correlation between $\gamma$-ray and far infrared (FIR) luminosities \citep{Rojas16}.  While \citet{Peng16} take this regularity to be an indication that the $\gamma$-rays in Arp~220 are solely the products of starburst activity, well-known Seyfert 2 galaxies also lie on this correlation, including NGC~1068, NGC~4945, and Circinus.  Cosmic ray interaction models for NGC~1068 demonstrate that $\gamma$-ray emission from cosmic rays accelerated by supernova remnants (SNRs) alone cannot reproduce the observed $\gamma$-ray flux \citep{Lenain10,YoastHull14a,Eichmann16}. Thus, activity from the central AGN of NGC~1068, or the associated radio jets, is most likely responsible for the observed $\gamma$-ray emission.

%
%
\begin{table*}
\begin{minipage}{170mm}
\begin{center}
\caption{Input Model Parameters}
\begin{tabular}{llllc}
\hline
Physical Parameters & Eastern Nucleus & Western ST & Western CND & References\\
\hline
CMZ Radius & 70 pc & 90 pc & 30 pc & 1,2,3\\
CMZ Disc Scale Height & 40 pc & 40 pc & 40 pc & 4\\
Molecular Gas Mass & $6 \times 10^{8}$ M$_{\odot}$ & $4 \times 10^{8}$ M$_{\odot}$ & $6 \times 10^{8}$ M$_{\odot}$ & 2,5 \\
Ionized Gas Mass & $3 \times 10^{6}$ M$_{\odot}$ & $2 \times 10^{6}$ M$_{\odot}$ & $3 \times 10^{6}$ M$_{\odot}$ & \\
Average ISM Density$^{a}$ & $\sim 7,700$ cm$^{-3}$ & $\sim 3,500$ cm$^{-3}$ & $\sim 42,000$ cm$^{-3}$ & \\
FIR Luminosity & $3\times 10^{11}$ L$_{\odot}$ & $3\times 10^{11}$ L$_{\odot}$ & $6\times 10^{11}$ L$_{\odot}$ & 2\\
FIR Radiation Field Energy Density$^{a}$ & 40,000 eV~cm$^{-3}$ & 27,000 eV~cm$^{-3}$ & 440,000 eV~cm$^{-3}$ & \\
Dust Temperature & 90 K & 50 K & 170 K & 2,6\\
Magnetic Field Strength & 2.0 mG & 1.0 mG & 3.5 mG & \\
\hline
\multicolumn{5}{l}{$^{a}$Derived from above parameters; References -- (1)~\cite{Downes07}; (2)~\cite{Sakamoto08};}\\ 
\multicolumn{5}{l}{ (3)~\cite{Aalto09}; (4)~\cite{Scoville15}; (5)~\cite{Downes98}; (6)~\cite{Wilson14}; }\\
\end{tabular}
\end{center}
\end{minipage}
\end{table*}

Prior to the detection of Arp 220 in $\gamma$-rays, \citet{Lacki13b} and \citet{YoastHull15} made predictions for the $\gamma$-ray flux based on cosmic ray interactions models constrained by the observed supernova rate and the radio spectra of the nuclei of Arp 220.  Unlike the estimates from the scaling relations in \citet{Griffin16} and \citet{Peng16}, \citet{Lacki13b} and \citet{YoastHull15} both calculated fluxes well below the observed $\gamma$-ray emission.

In this paper, we constrain the cosmic ray populations of the Arp 220 nuclei, and consequently the required star-formation rates (SFRs), from the observed $\gamma$-ray spectrum.  Utilizing an updated version of our semi-analytic cosmic ray interaction code \citep[][hereafter YEGZ]{YoastHull13}, we model the eastern nucleus as a single zone and the western nucleus as two spatial zones, an inner circumnuclear disc (CND) with a radius of $\sim 30$~pc and a surrounding torus with a radius of $\sim 90$~pc \citep{YoastHull15}, see Table 1.  $\chi^{2}$ tests allow us to explore the impacts of the total energy in cosmic rays and an energy-independent escape timescale on the $\gamma$-ray spectrum (Section 3.2) and the effects of ISM density and magnetic field strength on the radio spectrum (Section 3.3).  

\section{Properties of the Arp 220 Nuclei}

Arp~220 is a major merger of two disc galaxies and is the archetypal ULIRG \citep[e.g.,][]{Downes98}. Most of its bolometric luminosity, $L_{\text{bol}} \sim 10^{12}$ L$_{\odot}$, is emitted by its two nuclei. The high SFR of the two nuclei inferred from radio observations \citep[$\sim 200$~M$_{\odot}$~yr$^{-1}$, e.g.,][]{Barcos15,Varenius16} agrees with their FIR luminosity, assuming a standard relationship between the SFR and the FIR luminosity \citep{Kennicutt12}.  This demonstrates that a significant fraction of the power is produced by an intense starburst and that there is no clear evidence for AGNs in Arp~220 on the basis of energetics. However, the bolometric luminosities of the nuclei and their SFRs are uncertain \citep[e.g.,][]{Sakamoto08,Wilson14}. Therefore, the agreement between the SFRs and the luminosities of the nuclei does not exclude the presence of AGNs. 

Unfortunately the high dust opacities in Arp~220, particularly in the western nucleus, limit the utilization of standard indicators for the presence of AGN \citep[e.g.,][]{Teng15}. For example, the lack of observable Fe K-$\alpha$ emission does not exclude AGN (Paggi et al. 2017, in preparation) as the nuclei are Compton thick and X-rays are absorbed at the column densities observed in the western nucleus ($N_{H} \> 10^{25}$ cm$^{-2}$).  Molecular lines also are not reliable indicators of AGN activity as they can result from intense fluxes of cosmic rays and/or X-rays produced by a combination of supernovae and AGN \citep{Gonzalez2013}.  Additionally, the PAH features that are observed in infrared spectra likely originate from a lower density foreground than from the high opacity nuclei \citep{Soifer2002,Spoon2004}. 

Several arguments, however, suggest that AGNs exist in and provide significant fractions of the luminosity from Arp~220. \cite{Downes07} presented high angular resolution measurements at 1.3mm  that revealed a high brightness temperature in the core of the western nucleus. They interpreted their data in terms of a hot ($T \approx 100$~K), highly opaque disc that captured the luminosity from a hidden AGN. This model for the western nucleus is supported by observations of the dust content and molecular lines by the Atacama Large Millimeter Array \citep[ALMA;][]{Wilson14,Martin16}, which point to a warm and deeply embedded central source in the western nucleus. It also receives support from low frequency radio continuum observations with LOFAR \citep{Varenius16}.

The recent detection of Arp~220 as a luminous $\gamma$-ray source presents the opportunity to revisit the case for AGN in Arp~220. Our original model for the $\gamma$-ray luminosity from the starburst component underestimated the $\gamma$-ray luminosity by an order of magnitude \citep{YoastHull15}. Here we consider how varying the parameters for the Arp~220 starburst affects the $\gamma$-ray luminosity as a means to investigate whether the starburst can reasonably produce the observed $\gamma$-ray luminosity. 

\section{Model and Results}

\begin{figure*}
\centering
 \subfigure{
  \includegraphics[width=0.7\linewidth]{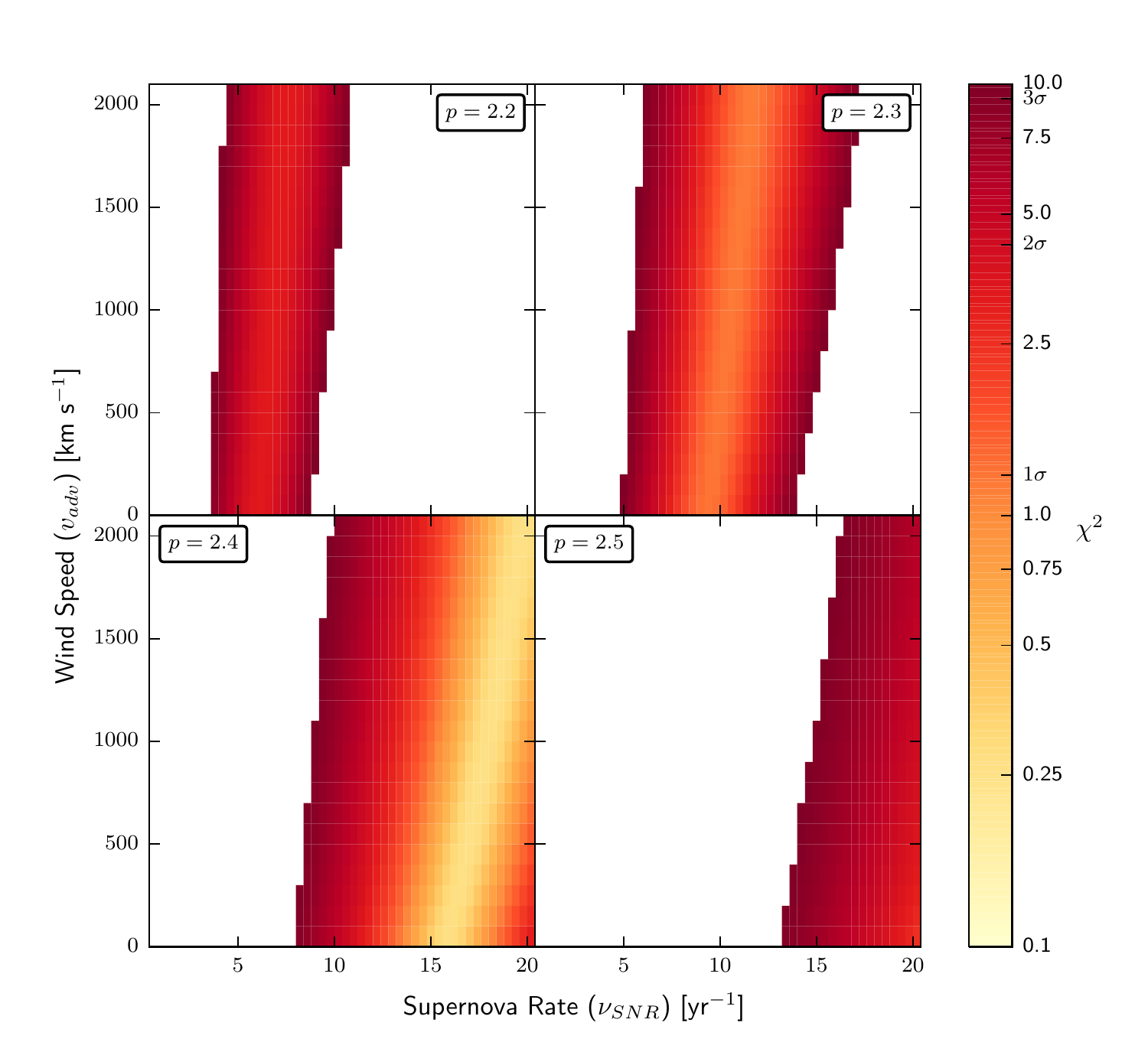}}
\caption{$\chi^{2}$ contour plots for the $\gamma$-rays.}
\end{figure*}
%
%

\subsection{Theoretical Model}

Beginning in YEGZ, we developed a semi-analytic, single zone model for cosmic ray interactions in the central molecular zones (CMZs) of starburst galaxies.  Assuming SNRs to be the primary accelerators of cosmic rays, up to PeV energies in Arp 220, we adopted a power-law injection spectrum directly proportional to the supernova rate and the fraction of the explosion energy transferred to cosmic rays.  The total cosmic ray spectrum is the product of the injection spectrum and the cosmic ray lifetime, which includes both energy losses and advective losses.  Energy losses include ionization, pion production, bremsstrahlung, inverse Compton, and synchrotron emission.

The YEGZ model accounts for secondary electrons and positrons resulting from the decay of charged pions.  The total $\gamma$-ray spectrum includes both hadronic (neutral pion decay) and leptonic (bremsstrahlung, inverse Compton) emission processes, and the radio spectrum includes non-thermal synchrotron emission as well as $\gamma-\gamma$ absorption and and free-free absorption and emission \citep{YoastHull14a}.

\subsection{Gamma-Ray Emission}

To model the nuclei of Arp 220, we adopted several of the same parameters used in our previous work \citep{YoastHull15}.  For the FIR luminosity, we assumed a ratio of 1 to 3 between the eastern and western nuclei such that $L_{\text{FIR}, E} = 3 \times 10^{11}$ L$_{\odot}$, $L_{\text{FIR}, W} = 9 \times 10^{11}$ L$_{\odot}$ \citep{Sakamoto08,Wilson14}.  This translates into the same ratio in the radiation energy densities ($U_{\text{rad}}$) and the supernova rate ($\nu_{\text{SN}}$, see Table 1).  For the molecular gas mass, we adopted a ratio of $\sim 1$ to 2 between the eastern and western nuclei such that $M_{\text{mol}, E} = 6 \times 10^{8}$ M$_{\odot}$, $M_{\text{mol}, W} = 1.0 \times 10^{9}$ M$_{\odot}$ \citep{Sakamoto08}.  Additionally, we adopted magnetic field strengths of $B_{E} = 2.0$~mG and $B_{W} = 1.0$~mG based on our previous best-fitting models \citep{YoastHull15}.

\begin{figure*}
\centering
 \subfigure[$\gamma$-Ray Spectrum]{
  \includegraphics[width=0.45\linewidth]{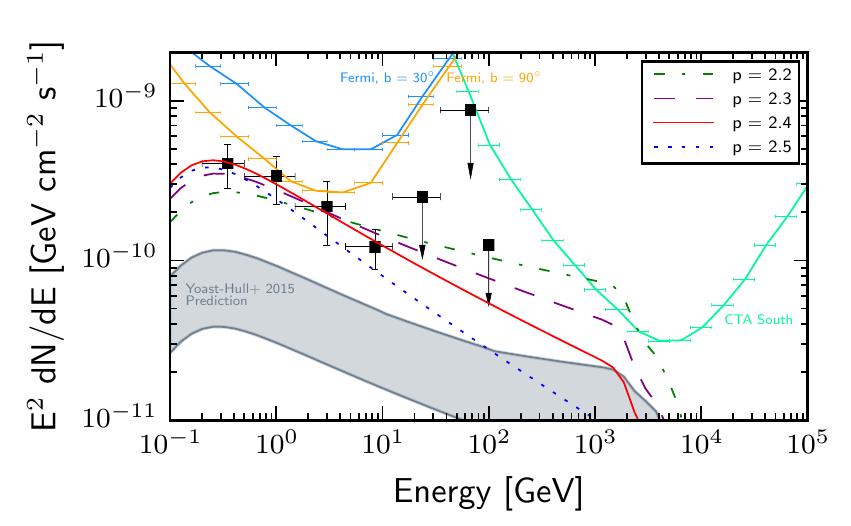}}
 \subfigure[Radio Spectrum]{
  \includegraphics[width=0.45\linewidth]{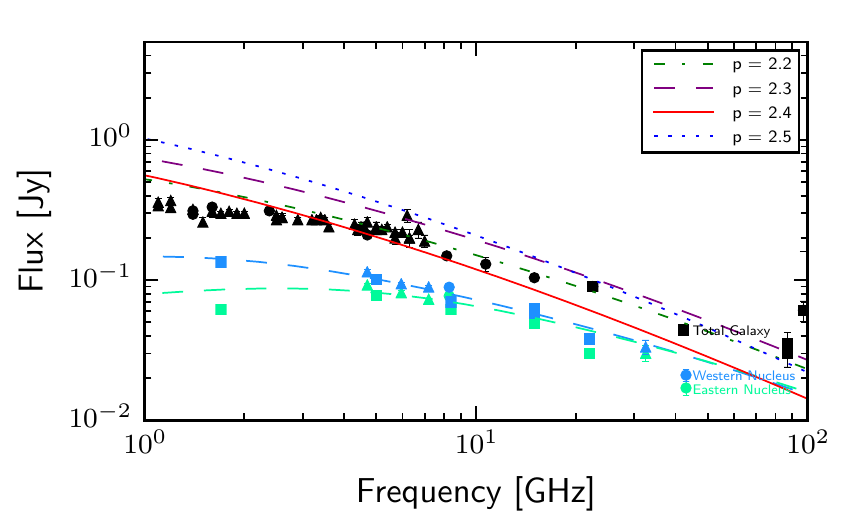}}
\caption{Plots of the total $\gamma$-ray and radio spectra from the YEGZ models.  Shown in both figures are our best-fitting $\gamma$-ray models and their corresponding radio models for various spectral indices: $p = 2.2$ (dot-dashed dark green line), $p = 2.3$ (dashed purple line), $p = 2.4$ (solid red line; best-fitting model), and $p = 2.5$ (dotted dark blue line).  These models assume molecular gas masses and magnetic field strengths from our previous fits to the radio spectrum, see Table 1 and \citet{YoastHull15}. In addition, new fits to the radio spectrum are included for the eastern (light green) and western (light blue) nuclei, assuming $\nu_{\text{SN}} = 15.9$ yr$^{-1}$ and $v_{\text{adv}} = 0$ km~s$^{-1}$, and our previous estimates for the $\gamma$-ray spectrum from \citet{YoastHull15} are included in gray and sensitivity limits are shown for both \textit{Fermi} (solid light blue and orange lines) and CTA (solid light green line).  Observational data are taken from \citet{Peng16}, \citet{Barcos15}, \citet{Williams10}, \citet{Rodriguez05}, and \citet{Downes98}.} 
\end{figure*}

To best constrain the total energy input into cosmic rays necessary to reproduce the observed $\gamma$-ray flux, we varied the supernova rate and the advective wind speed, and we used $\chi^{2}$ tests to determine the best-fitting parameters.  We tested wind speeds from $v_{\text{adv}} = 0$ to 2000 km~s$^{-1}$, supernova rates from $\nu_{\text{SN}} = 0.4$ to 20 yr$^{-1}$, and spectral indices from $p = 2.2 - 2.5$ (see Fig. 1).  The best-fitting models occur for $p = 2.4$, and we also found models within $1\sigma$ of the minimum $\chi^{2}$ value for $p = 2.3$ but no acceptable models for either $p = 2.2$ or $p = 2.5$ (see Fig. 1 \& 2).

We observed a degeneracy between the wind speed and the supernova rate for the best-fitting models such that as the supernova rate increases, the wind speed must also increase to achieve the same goodness of fit.  This is expected as increasing the wind speed is equivalent to decreasing the total cosmic-ray proton population and increasing the supernova rate is equivalent to increasing the total cosmic-ray population.  Thus, the best-fitting models ($\chi^{2} \sim 0.25$) have supernova rates ranging from $\nu_{\text{SN}} = 15.6 - 19.6$~yr$^{-1}$.  These rates are well above the observed supernova rate of $\sim 2 - 4$~yr$^{-1}$ \citep{Varenius16}, indicating that another energy source is likely necessary to explain the observed $\gamma$-ray flux.

\subsection{Radio Emission}

Taking the best-fitting results from our analysis of the $\gamma$-ray spectrum, we found that the resulting radio spectra are more luminous than the observed data by factors of $\sim 5 - 10$ (see Fig. 2).  Additionally, the resulting radio fluxes exceed those observed for the entire galaxy.  This feature of the flux the models follows naturally from the assumption of very high supernova rates.

Of course, the calculated radio flux depends significantly on the assumed electron to proton ratio for the primary cosmic rays ($\xi$), the assumed magnetic field strength ($B$), and the average ISM density ($n_{\text{ISM}}$).  To determine whether the radio spectrum could be better reproduced, we adopted a wider parameter space.  We explored the effects of changes in the magnetic field strength and average density, leaving the assumed ratio of primary electrons to protons at $\xi = Q_{e} / Q_{p} = 0.02$.

\begin{figure}
\centering
 \subfigure{
  \includegraphics[width=0.8\linewidth]{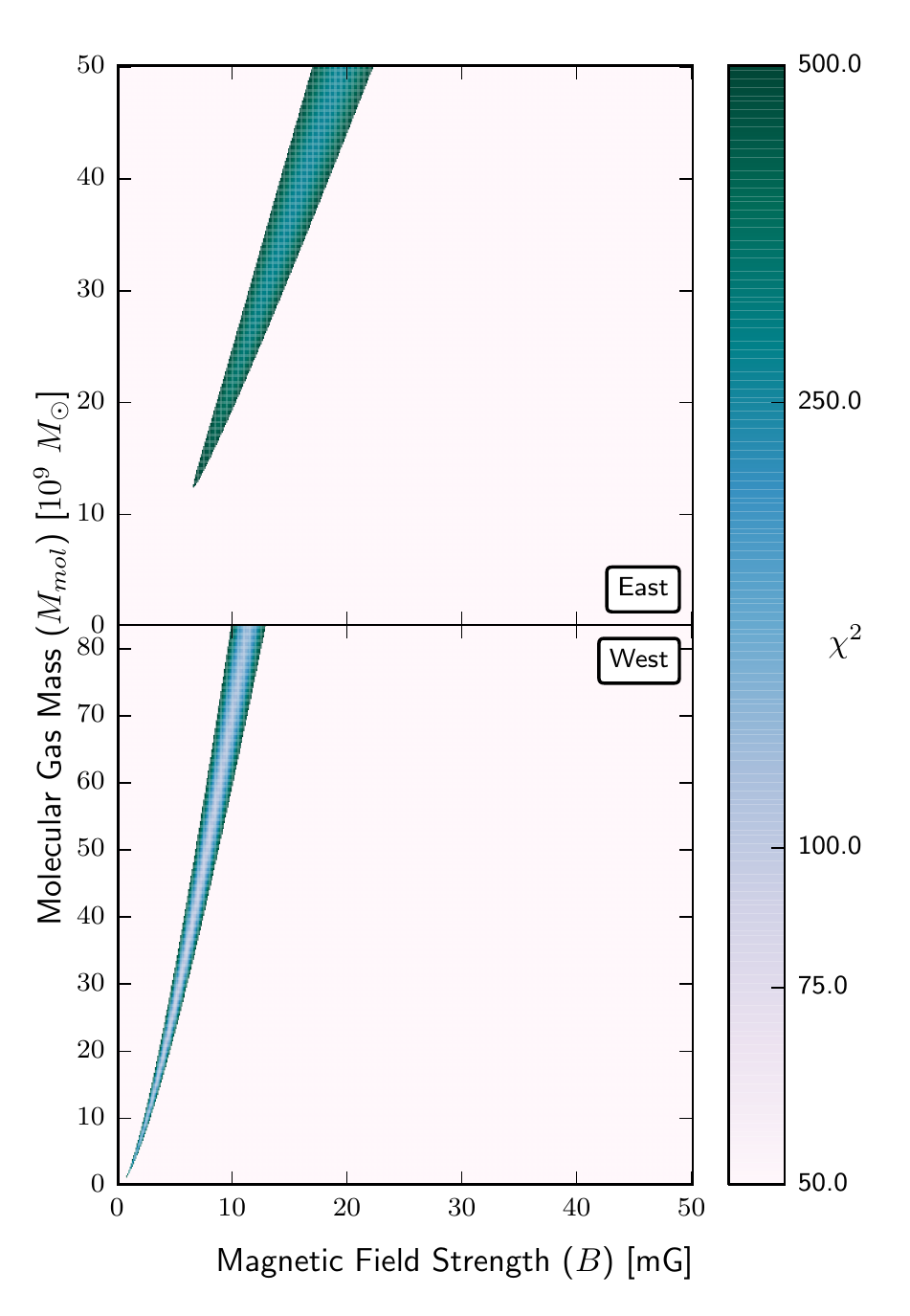}}
\caption{$\chi^{2}$ contour plots for the radio.}
\end{figure}

Varying magnetic field strength from $B = 0.1 - 50.0$ mG and the molecular gas mass from $M_{\text{mol},E} = (0.1 - 50) \times 10^{9}$ M$_{\odot}$, we used $\chi^{2}$ tests to compare the modeled radio spectrum with the observed data (see Fig. 3).  We tested these models in the calorimeter limit (no advection, $v_{\text{adv}} = 0$ km~s$^{-1}$) to reduce computation time, as the average ISM density then cancels out of the equations for the secondary electron and positron populations.  The average density remains as a factor only in the electron energy-loss lifetimes.

Comparing the non-thermal synchrotron emission with the observed radio data, we find that the models closest to fitting the radio spectra occur for $B_{E} = 16.9$ mG and $M_{\text{mol},E} = 4.2 \times 10^{10}$ M$_{\odot}$ in the eastern nucleus and $B_{W} = 6.3$ mG and $M_{\text{mol},W} = 3.7 \times 10^{10}$ M$_{\odot}$ in the western nucleus (see Fig. 3).  Significant increases in ISM density result in bremsstrahlung becoming the dominant energy loss mechanism for electrons at near GeV energies.  This translates into a flattening of the cosmic-ray electron lifetime and thus a significant flattening of the radio spectrum at low frequencies (YEGZ).  While this effect is seen in our models, see Fig. 2(b), a flattening of the radio spectrum at low frequencies could also be attributed to free-free absorption.  However, the ionized gas contents of the Arp 220 nuclei are highly uncertain; thus, we did not include their effects here.

Measurements of the magnetic field via Zeeman splitting give strengths of $\sim 1 - 5$ mG in the nuclear regions of Arp 220 \citep{McBride15}.  Thus, while our results for the western nucleus are nearly in agreement with observations, our results for the eastern nucleus result in magnetic field strengths roughly a factor of $5 - 10$ larger than observations.  Additionally, the necessary molecular gas masses correspond to column densities on the order of $N_{ISM} \sim 10^{26}$ cm$^{-2}$ which is roughly an order of magnitude above current measurements and estimates \citep{Wilson14,Martin16}.  Thus, our `best-fitting' radio models are not compatible with either the radio data or current ISM observations.

\section{Discussion and Summary}
Our YEGZ model for radio and $\gamma$-ray emission from starburst CMZs makes standard assumptions about the astrophysics, fits nearby starbursts, and agrees in its essentials with other modeling approaches. Our best-fitting models for the $\gamma$-ray spectrum of Arp~220 require that $\nu_{\text{SN}} \geq 15$~yr$^{-1}$. This result disagrees with the observed $\nu_{\text{SN}} \sim 2 - 4$~yr$^{-1}$ and leads to a major overestimate of both the bolometric and radio luminosities. Thus, a normal starburst model {\it does not} reproduce the $\gamma$-ray luminosity of Arp~220.

A buried AGN is a plausible explanation for the $\gamma$-ray luminosity observed in Arp 220.  For instance, we know that in some moderately obscured systems, AGN can produce substantial $\gamma$-ray luminosities, such as that in the Seyfert~2 galaxy NGC~1068.  AGN models can produce energetic $\gamma$-rays from inverse Compton (IC) radiation resulting from the interaction of highly relativistic e$^{+/-}$ with synchrotron or other radiation fields. Additional possibilities exist for cosmic ray and $\gamma$-ray production, including acceleration of protons by AGN jet shocks or relativistic proton production near SMBHs \citep[e.g.,][]{Oka2003,Stecker2013}. In Arp 220, we find an AGN which is a powerful $\gamma$-ray source with a weak radio synchrotron component.  This could be achieved by a mildly relativistic jet where e$^{+/-}$ energy losses are dominated by bremsstrahlung and IC off the intense far-infrared radiation field.

Alternative options to AGNs would require the presence of an unusual source of energetic cosmic rays that is not tied to particle acceleration in typical SNRs.  While such circumstances should not be quickly or simply excluded in the complex and extreme Arp~220 system, the nature of such a mechanism is not known, and there are no clear hints for such a presence in radio maps which trace synchrotron emission \citep{Barcos15,Varenius16}. 

We conclude that the present data are best understood by the presence of a highly obscured AGN in the western nucleus of Arp~220. The western nucleus is the preferred location for the main AGN in Arp~220 due to its high infrared luminosity and incredible optical depth which is most simply understood if the energy source is very compact \citep[e.g.,][]{Downes07,Martin16}. Unfortunately given the uncertainties about the origin of $\gamma$-rays in AGNs, we cannot directly estimate the luminosity of the AGN in the western nucleus. However, an AGN producing a reasonable fraction of the bolometric luminosity of the western nucleus (e.g., $L_{\text{bol}} \sim 10^{11.3}$ L$_{\odot}$) would have a SMBH with $M \sim 3 \times 10^7$ M$_{\odot}$ accreting at an Eddington rate of somewhat less than 1 M$_{\odot}$~yr$^{-1}$. These parameters are reasonable for the SMBH in a massive spiral.

The likely existence of an AGN in the western nucleus of Arp~220 has interesting implications in that the SMBH could be experiencing significant growth. Given the current SFR of $\sim 200$ M$_{\odot}$~yr$^{-1}$ in the circumnuclear molecular disc associated with the western nucleus, the gas reservoir  will last for less than about $10^{7}$~yr. During such a short time span, the SMBH only would experience modest growth at most if nothing in the system were to change. 

The optical depth of the west nucleus region evidently has hidden every indicator of AGN activity aside from the $\gamma$-ray luminosity. This is especially impressive as the main molecular disc of the west nucleus is only moderately inclined, suggesting that the deep obscuration occurs close to the AGN. If this type of behavior also occurs at high redshifts, then the presence of powerful AGN contributing to the luminosities of dusty galaxies could be completely hidden from the usual mid-infrared and X-ray diagnostics. 

\section*{Acknowledgements}

This work was supported in part by NSF AST-0907837 and UW-Madison.  We thank Ellen Zweibel for useful discussions.

%
\end{document}